\documentclass[12pt]{article}
\usepackage[cp1251]{inputenc}
\usepackage[english,russian]{babel}
\usepackage{graphicx}
\usepackage{mathtext}
\usepackage{epsfig,amsmath,amsfonts}
\usepackage{hyperref}

\inputencoding{cp1251}

\def\a{\alpha}   \def\d{\delta}

 \def\s{\sigma}   
\def\ff{\phi}

   \def\L{\Lambda} 
  \def\P{\Pi}

\def\fr{\frac}  \def\dt{\partial}

\def\beq{\begin{equation}}
\def\eeq{\end{equation}}

\def\bear{\begin{eqnarray}}
\def\eear{\end{eqnarray}}

\def\bea*{\begin{eqnarray*}}
\def\eea*{\end{eqnarray*}}

\def\mc{\mathcal}

\def\Tr{\mbox{Tr}}

\begin{document}

\renewcommand{\contentsname}{}
\renewcommand{\refname}{\begin{center}References\end{center}}
\renewcommand{\abstractname}{\begin{center}Abstract\end{center}}

\begin{center}
{\Large\bf Hints on integrability in the \\[.2cm]Wilsonian/holographic  renormalization group}
\\[1.5cm]
{\large E.T. Akhmedov$^{\,*\,\ddagger}$, I.B. Gahramanov$^{\,\dagger}$,
E.T. Musaev$^{\,*\,\ddagger}$}\\[6mm]
{\it
$^{*}$~Moscow Institute of Physics and Technology, Dolgoprudny, Russia\\[3mm]
$^{\dagger}$~National University of Science and Technology ``MISIS'', Moscow, Russia\\[.3cm]
$^{\ddagger}$~Institute for Theoretical and Experimental Physics, B.Cheremushkinskaya, 25, 117218, Moscow, Russia} \\[.5cm]
{\small \href{mailto:akhmedov@itep.ru}{\tt akhmedov@itep.ru}, \href{mailto:ilmargh@gmail.com}{\tt ilmar.gh@gmail.com}, \href{mailto:musaev@itep.ru}{\tt musaev@itep.ru}}
\\[2.0cm]
\end{center}

{\abstract 
The Polchinski equations for the Wilsonian renormalization group in the $D$--dimensional matrix scalar field theory can be written at large $N$ in a Hamiltonian form. The Hamiltonian defines evolution along one extra holographic dimension (energy scale) and can be found exactly for the complete basis of single trace operators.  We show that at low energies independently of the dimensionality $D$ the Hamiltonian system in question (for the subsector of operators without derivatives) reduces to the {\it integrable} effective theory. The obtained Hamiltonian system describes large wavelength KdV type (Burger--Hopf) equation and is related to the effective theory obtained by Das and Jevicki for the matrix quantum mechanics.}

\vspace{10mm}

{\bf Introduction.}
One of the greatest achievements of the modern fundamental physics is the holographic duality between $D$--dimensional gauge and $(D+1)$--dimensional gravity theories. The seminal example of this duality is the equality between the quantum generating functional of the correlation functions for the flat space gauge theory and the classical wave functional for the AdS gravity theory \cite{Maldacena:1997re}. The latter is known as AdS/CFT--correspondence.

The extra dimension in the gravity theory has a natural interpretation as the energy scale in the gauge theory \cite{Maldacena:1997re} (see e.g. \cite{Akhmedov:2009zz} for a review). Moreover, $(D+1)$--dimensional gravity equations of motion can be related to the renormalization group (RG) equations on the gauge theory side \cite{Akhmedov:1998vf} (see as well \cite{Balasubramanian:1999jd}--\cite{Akhm}).

To understand deeper such a holographic duality, we would like to address the following  more general question: what kind of the $(D+1)$--dimensional theories govern RG flows of the large $N$ $D$--dimensional field theories? In \cite{Akhmedov:2010sw} matrix scalar field theory was considered. The Polchinski \cite{Polchinski:1983gv} equations for the Wilsonian RG for this theory were formulated. At large $N$ the Polchinski equations reduce to the Hamiltonian ones.
The latter Hamiltonian system is rather artificial and contains non-local terms. However, in this note we show that in the infrared (IR) limit the RG dynamics is governed by the Hamiltonian\footnote{Below in this paper we correct the important mistake made by two of us in the paper \cite{Akhmedov:2010sw}. As well here we generalize \cite{Akhmedov:2010sw} and write the RG equations in the Hamiltonian form for the complete basis of single trace operators, including those which contain derivatives.} \cite{Akhmedov:2010sw}:

\begin{equation}
\label{00.1}
 H=\int_{-\pi}^{+\pi}d\s \int d^Dx \, \P^2J'.
\end{equation}
Here $J'= dJ/d\s$,
$J(T,\s,x) = \sum_k \s^{k}\,J_k(T,x)$, $J_k(T,x)$ are sources only for the operators without derivatives --- $\Tr[\phi^k(x)]$;
$\P(T,\s,x) = \sum_k \s^{-(k+1)}\,\P_k(T,x)$, $\P_k(T,x)$ are vaguely speaking vacuum expectation values (VEV) of the operators $\Tr[ \phi^k(x)]$ (see below). The sources and VEVs are conjugate to each other via the Poisson bracket: $\left\{\P(T,\s,x)^{\phantom{\frac12}},\,\, J(T,\s',x')\right\} \propto \delta(\s-\s')\,\delta(x-x')$.
On the $D$--dimensional matrix scalar field theory side this can be traced from the Legendre (functional Fourier) relation between the effective actions for the sources and VEVs \cite{Akhmedov:2010sw}. The role of the time $T$ for the Hamiltonian system in question is played by the energy scale in the scalar field theory. Note that we observe here the appearance of the one extra (on top of the energy scale) dimension $\s$ conjugate to the number $k$ enumerating the operators $\Tr[\phi^k(x)]$.

Usually in the holographic duality the $(D+1)$--dimensional theory contains gravity. The low energy Hamiltonian system (\ref{00.1}) does not contain the symmetric tensor particle because among the Tr$\phi^l(x)$ operators there is no energy--momentum tensor of the matrix field theory. Hence, naively there is no gravity in the above theory. However, the full Hamiltonian driving the RG flow on the complete basis of single trace operators contains symmetric tensor particle with the appropriate number of degrees of freedom.

In general, however, it is not clear for us so far under what circumstances and/or how
the Hamiltonian equations in question are converted into the Hamiltonian constraint equations of the generally covariant theory. It is not clear for us whether the theory which governs the RG flow of field theory should always (for all large $N$ field theories) be generally covariant or not. This remains to be a challenge for the future work.
However, we refer to the RG in question as holographic because, unlike the standard definition of the RG, one can find the theory at any energy scale (high or low one) once he specified this theory at some scale. As well we believe that AdS/CFT--correspondence is of the same origin.

One of the goals of this paper is to show that the Hamiltonian system (\ref{00.1}) is integrable.
And to show that it is equivalent to the effective field theory derived by Das and Jevicki for the matrix quantum mechanics \cite{Das:1990kaa} (see as well  \cite{Jevicki:1993qn},\cite{Polchinski:1994mb}).

{\bf Holographic formulation of the large $N$ Wilsonian RG.}
We consider the $D$--dimensional Euclidian matrix scalar field theory whose action is

\begin{multline}
\mc{S}[\ff]=-\fr{N}{2}\int \Tr \partial_\mu \phi(x)\, \partial_\mu \phi(x)\, d^D x + \\ + N\sum_{l=0}^{\infty}\int d^D x_1\ldots{}d^D x_l\,\Tr \left[\ff(x_1)^{\phantom{\frac12}}\ldots\,\ff(x_l)\right]\,J_l(x_1,\ldots,x_l).
\end{multline}
Here $J_l$ are the sources. Note that we include only single trace operators. The second term in this action can be understood as shorthand notation for writing sources for the complete basis of single trace operators with derivatives.

Regularized action in the Fourier transformed form can be written as:

\begin{multline}
 \label{1.1}
\mc{S}[\ff]=-\fr{N}{2}\int \Tr\left[\ff(p)\,(p^2+m^2)^{\phantom{\frac12}}K_{\L}^{-1}(p^2)\,\ff(-p)\right]\, d^D p + \\ + N\sum_{l=0}^{\infty}\int d^D k_1\ldots{}d^D k_l\,\Tr \left[\ff(k_1)^{\phantom{\frac12}}\ldots\,\ff(k_l)\right]\,J_l(k_1,\ldots, k_l).
\end{multline}
We assume that there is some momentum cut--off imposed, i.e. $K_\Lambda(p^2)\sim 1$ as $p^2<<\L^2$, while $K_\Lambda(p^2) \rightarrow 0$ as $p^2>>\L^2$. As well we assume that $J_l(k_1, \dots, k_l)=0$ for all $l$ and for all $|k_l|>\lambda$, where $\lambda$ is some low energy scale (where we measure our physics).

The Polchinski equation for the theory in question follows from the RG invariance of the functional integral $Z$ \cite{Polchinski:1983gv}:

\begin{eqnarray}
\label{1.2}
\L\fr{d\mc{S}_I[\ff]}{d\L} = -\fr{1}{2}\int\fr{d^D p}{p^2+m^2} \, \L\, \fr{dK_{\L}(p^2)}{d\L} \, \left[N^{-1}\fr{\d^2\mc{S}_I[\ff]}{\d\ff^{ij}(-p)\d\ff^{ji}(p)} + \fr{\d\mc{S}_I[\ff]}{\d\ff^{ij}(p)}\fr{\d\mc{S}_I[\ff]}{\d\ff^{ji}(-p)}\right],\nonumber\\
\mc{S}_{I}=\sum_{l=0}^{\infty}\int d^D k_1\ldots{}d^D k_l\,\Tr \left[\ff(k_1)^{\phantom{\frac12}}\ldots\,\ff(k_l)\right]\,J_l(k_1,\ldots,k_l),
\end{eqnarray}
i.e. this equation is supposed to specify the scale $\Lambda$ dependence of the sources $J_l$ to fulfill the equation $dZ/d\Lambda=0$. On the RHS of this Polchinski equation there is an extra term which contributes only to the RG dynamics of the unit operator. We drop this term and do not consider unit operator below, because its dynamics decouples from the rest of the system --- it does not participate into the RG equations for the other operators.

To proceed, we verify the following relations:

\begin{eqnarray}
\Tr\left[\fr{\d^2\mc{S}_I}{\d\ff(p)\d\ff(-p)}\right] = \sum_{l=1}^{\infty} \sum_{a,b=1}^{l} \, \int dk_1\dots \hat{dk_a}\dots \hat{dk_b}\dots dk_l  \times \nonumber \\ \times  \Tr\left[\ff(k_{a+1})^{\phantom{\frac12}} \ldots \, \ff(k_{b-1})\right]\, \Tr\left[\ff(k_{b+1})^{\phantom{\frac12}} \ldots \,\ff(k_{a-1})\right]\times \nonumber \\ \times J_l\left(p,k_{a+1},\dots, k_{b-1}, -p, k_{b+1}, \dots, k_{a-1}\right);\nonumber \\
\Tr\left[\fr{\d \mc{S}_I}{\d\ff(p)}\fr{\d \mc{S}_I}{\d\ff(-p)}\right] = \sum_{l,j=1}^{\infty}\sum_{a=1}^l \sum_{b=1}^j \int dq_1\dots \hat{dq_a}\dots dq_l dk_1\dots \hat{dk_b}\dots dk_j \times \nonumber \\ \times \Tr\left[\ff(q_1) \ldots \hat{\phi(q_a)}\dots \ff(q_{l})^{\phantom{\frac12}}\ff(k_1)\ldots \hat{\phi(k_b)} \dots \ff(k_j)\right]\times \nonumber \\
\times{} J_l\left(p, q_{a+1},\ldots, q_{a-1}\right)\,J_j\left(-p, k_{b+1},\ldots, k_{b-1}\right).
\end{eqnarray}
Hat over the quantity means that we omit it in the expression. In deriving these expressions we have used the cyclic symmetry of the sources $J_l(p_1,\dots, p_l)$. To shorten the formulas we will use the notations

\begin{eqnarray}
\label{not}
\int dk_1\dots \hat{dk_a}\dots \hat{dk_b}\dots dk_l := \int dk_{1-a-b-l},
\nonumber \\
\int dq_1\dots \hat{dq_a}\dots dq_l dk_1\dots \hat{dk_b}\dots dk_j := \int dq_{1-a-l} \int dk_{1-b-j},
\nonumber \\
J_l\left(p,k_{a+1},\dots, k_{b-1}, -p, k_{b+1}, \dots, k_{a-1}\right) := J_l\left(p, k_{a-b}, -p , k_{b-a}\right), \nonumber \\
J_l\left(p, q_{a+1},\ldots, q_{a-1}\right) := J_l\left(p, q_{a-a}\right).
\end{eqnarray}
Note that $a$ in the sums can be grater than $b$.

If one substitutes the obtained expressions for the variations of $\mc{S}_I$ into (\ref{1.2}), he encounters higher trace operators on its RHS.
It seems that to close the obtained system one has to add to (\ref{1.1}) sources for higher trace operators as well. That is the standard approach in the Wilsonian RG: at the end one has to use Operator Product Expansion and the completeness of the basis of operators.

However, in the large $N$ limit one can take a different way of addressing the problem \cite{Akhm}.
In this limit one has the factorization property, $\langle \prod_n O_n \rangle = \prod_n \langle O_n \rangle + {\cal O}\left(1/N^2\right)$, which is valid for any choice of the bare action used for performing the quantum average $\langle \dots \rangle$. Due to the factorization property at large $N$ one can express any operator from the OPE algebra as a polynomial in the single trace operators. Saying another way, in the limit in question single trace operators form a basis through which any operator in the complete OPE algebra can be expressed algebraically.

Thus, the first step to obtain the closed system of equations for the single trace operators is to take the quantum average of (\ref{1.2}) \cite{Akhmedov:2010sw}.
To do that we separate the field $\phi$ into two contributions $\phi=\phi_0+\varphi$ --- the low energy one $\phi_0$, which solves the equations of motion following from (\ref{1.1})\footnote{This field contains harmonics lower than the low energy scale $\lambda$, because as we assume the sources $J_l(k_1,\dots, k_l)$ are zero when the moduli of their arguments are grater than $\lambda$.}, and the high energy ones $\varphi$, which contain harmonics between $\lambda$ and $\Lambda$. In the quantum average we take the functional integral over the $\varphi$ with the use of the full interacting action (\ref{1.1}).

Thus, substituting $\mc{S}_I$ to the equation \eqref{1.2}, then taking its quantum average and using the factorization property, one obtains:

\begin{eqnarray}
 \label{1.3}
\sum_{l=1}^{\infty}\int dk_{1-l} \, T_{l}(k_{1-l})\dot{J}_l(k_{1-l}) =
- \fr{1}{2}\int\fr{dp\,\dot{K}_\L(p^2)}{p^2+m^2} \times \nonumber \\ \times \left[ N^{-1}\, \sum_{l=1}^{\infty} \sum_{a,b=1}^{l} \int dk_{1-a-b-l}T_{|b-a|}(k_{a-b}) \, T_{l-|a-b|}(k_{b-a}) \, J_l(p,k_{a-b}, -p , k_{b-a})  \right.\nonumber \\ \left. + \sum_{l,j=1}^{\infty}\sum_{a=1}^l\sum_{b=1}^j \int dq_{1-a-l} \int dk_{1-b-j} \, T_{l+j-2}(k_{1-a-l},q_{1-b-j})\,J_l\left(p,q_{a-a}\right)\,J_j\left(-p,q_{b-b}\right)\right],
\end{eqnarray}
where overdot means $\Lambda\,d/d\Lambda$ and $T_l(k_{1-l}):=\langle\Tr[\ff(k_1)\ldots\ff(k_l)]\rangle$
and for $T$'s we use similar shorthand notations as (\ref{not}) for $J$'s. 

The equation (\ref{1.3}) still is not closed since the RG dynamics for the sources $J$ depends on the VEVs $T$. However, one can close the system by deriving the RG equations for the VEVs $T$ as well.
To derive the above Polchinski equation, we have used the fact that the effective action $W(J)=\log Z$
is cutoff independent. Because $W(J)$ is the effective action the VEV defined in the previous paragraph --- $T_l(k_1,\dots, k_l) = \delta W(J)/\delta J_l(k_1,\dots, k_l)$ --- is just the momentum conjugate to the source $J_l(k_1,\dots, k_l)$.
Then, we can make the Legendre transform from $W(J)$ to the effective action $I(T) = \left.\left[\int T\, J - W(J)\right]\right|_{T=\delta W/\delta J}$. The latter should not depend on the cutoff as well. Hence, from the RG invariance of $I(T)$ we expect to get a Hamiltonian conjugate equation describing the RG dynamics of $T$. That is explicitly checked in perturbation theory in \cite{Akhm}.

Now putting all terms in (\ref{1.3}) to the same side (and doing the same for the corresponding Hamiltonian conjugate one for $T$'s), then equating to zero each term in front of every operator $T_l$ in (\ref{1.3}) (and in front of every $J_l$ in the Hamiltonian conjugate equation), we obtain the system of equations of the Hamiltonian form $\dot{J}_l (k_{1-l}) = \delta H(J,T)/\delta T_l(k_{1-l})$ and $\dot{T}_l (k_{1-l}) = - \delta H(J,T)/\delta J_l(k_{1-l})$, where

\begin{eqnarray}
\label{genham}
H = - \fr{1}{2}\int\fr{dp\,\dot{K}_\L(p^2)}{p^2+m^2} \, \left[ N^{-1}\, \sum_{l=1}^{\infty} \sum_{a,b=1}^{l} \int dk_{1-a-b-l}T_{|b-a|}(k_{a-b}) \, T_{l-|a-b|}(k_{b-a}) \, J_l(p,k_{a-b}, -p , k_{b-a})  \right.\nonumber \\ \left. + \sum_{l,j=1}^{\infty}\sum_{a=1}^l\sum_{b=1}^j \int dq_{1-a-l} \int dk_{1-b-j} \, T_{l+j-2}(k_{1-a-l},q_{1-b-j})\,J_l\left(p,q_{a-a}\right)\,J_j\left(-p,q_{b-b}\right)\right].
\end{eqnarray}
as follows from (\ref{1.3}).
Such an approach within perturbation theory was verified in \cite{Akhm}. So far we have been making identity transformations (up to the derivation of the equations for $T$).

The ``Hamiltonian system'' under consideration, although being closed, is rather artificial at least because it does not have fixed dimensionality. All the terms in (\ref{1.3}) are relevant in the {\it UV limit} ($\lambda \sim \Lambda$). Which is necessary to restore the proper $\beta$--functions of the sources.
However, we are going to argue that in the {\it IR limit} ($\lambda << \Lambda$)
the second term on the RHS of (\ref{1.3}) is irrelevant.

Because the theory in question has Landau pole all $J$'s (even the ones corresponding to the marginal operators) scale to zero in IR limit under RG flow. Thus, in the IR limit the second term $\sim J^2$ is suppressed in comparison with the first term $\sim J$ and the RG dynamics is governed by the Hamiltonian (\ref{genham}) without the second term. Furthermore, in IR limit we should keep in the first term in (\ref{genham}) only those $J_l$ which are the sources for the operators without derivatives, i.e. only sources for Tr$\phi^l(x)$. Because Fourier transformed form of $\int d^Dx J_l(x)\, \Tr \phi^l(x)$ is $\int d^D k_1\ldots{}d^D k_l\,\Tr \left[\ff(k_1)^{\phantom{\frac12}}\ldots\,\ff(k_l)\right]\,J_l(-k_1 \ldots - k_l)$ we have to keep among $J_l(k_1,\dots, k_l)$ only those which depend on the sum of $k_l$'s rather than on all $k_l$ separately\footnote{The observations of the last two paragraphs correct the mistake which was made by two of us in the earlier paper \cite{Akhmedov:2010sw}, where we have represented the second term of the above ``Hamiltonian system'' in the ultra--local form.}.

Thus, if we neglect the second term in the IR limit and keep only $J_l(-k_1-\dots -k_l)$ sources the Hamiltonian in question reduces to:

\begin{equation}
\label{2}
H=\int dq_1\, dq_2 \,\sum_{l,s=0}^{\infty}\left[(l+s+2)\P_l(q_1)\P_s(q_2)J_{l+s+2}(-q_1-q_2)\right].
\end{equation}
The time in this theory is related to the scale factor as follows:
$T=\int \fr{dp \, K_\L(p^2)}{p^2+m^2}$.
The canonical momenta $\P_n(p)$ are now as follows:
$\P_n(p)=\frac{1}{N}\,\int_{p_1\ldots p_n}\d(p-p_1-\cdots-p_n)\langle\Tr[\ff(p_1)\ldots\ff(p_n)]\rangle$, where the average is taken over the high--energy harmonics $\varphi$.

To represent the Hamiltonian (\ref{2}) in the ultra--local form we introduce the Fourier transform of the $J_k$ and $\P_k$ harmonics: $J(T,\s,x)=\sum_k \s^k J_k(T,x), \quad \P(T,\s,x)=\sum_k \s^{-(k+1)}\P_k(T,x)$. After such a substitution the Hamiltonian acquires the simple form:

\begin{equation}
\label{H}
H=\int_{-\pi}^{\pi}d\s\, \int d^Dx \,\P^2J'.
\end{equation}
The dynamics of this Hamiltonian system along the $D$ directions (denoted by $x$) is trivial and we can skip the $x$ dependence of $J$ and $\P$.
Then the corresponding equations of motion are: $\dot{J}=2\P J', \quad \dot{\P}=2\P\P'$.
The equations of motion for the field $J(t,\s)$ have the form:

\begin{equation}
 \label{10}
 -\dt_t\left(\fr{\dot{J}}{J'}\right)+\frac12\dt_\s\left(\fr{\dot{J}}{J'}\right)^2=0.
\end{equation}
This equation (for $P=\dot{J}/J'$) is referred to as the inviscid Burger's or Hopf equation and is know to be integrable.

{\bf RG, matrix models, effective theory and strings.}
In the IR limit (when we can neglect the second term in (\ref{genham})) in the coordinate representation the Polchinski equation \eqref{1.3} can be written as

\begin{equation}
 \label{8}
\int d^Dx \sum_k\P_k(x)\dot{J}_k(x)= - \int d^Dx \sum_{k,l}(k+l+2)\P_k(x)\P_l(x)J_{k+l+2}(x),
\end{equation}
where overdot means the differentiation with respect to the above defined ``time'' $T$.

To establish a correspondence of the holographic RG equations and effective field theory of Das and Jevicki we redefine the sources and the conjugated momenta in Polchinski equation according to their natural conformal dimensions: $J_k(x)=\L^{\a k}g_k, \P_k(x)=\L^{-\a k}p_k$, where $\a=(2-D)/2$. Then the above Hamiltonian transforms to:

\begin{equation}
H = \left[\sum_{k,l}(k+l+2)\,p_k \,p_l \,g_{k+l+2}+\sum_k{k} p_k\,g_k\right].
\end{equation}
Here we have a different definition of time $t=-\fr{D-2}{2}\log\L$. The case of $D=2$ is special: one should take $\a=-1/2\,$ and $\P_k(x)=\L^{-k/2+1}\,p_k(x)$. The time then becomes $t_{D=2}=-1/2\log\L$ and the Hamiltonian is the same.

Making the Fourier transform as above and introducing the new variable $s=e^{i\s}$, we obtain:

\begin{equation}
 H=\int ds \left[p^2\,g'+s \,p \,g'\right].
\end{equation}
The equation of motion for the field $g(t,s)$ is as follows:

\begin{equation}
 -s-\dt_t\left(\fr{\dot{g}}{g'}\right)+\frac12\dt_s\left(\fr{\dot{g}}{g'}\right)^2=0.
\end{equation}
With the obvious change of variables $P=\dot{g}/g'$ this equation can be rewritten in a rather simple way:

\begin{equation}
 \dot{P}=P\,\dt_s\,P - s.
\end{equation}
Such an equation follows from the Hamiltonian system, which in the standard notations looks as

\begin{eqnarray}
H = \frac{1}{4\pi} \int ds \left[\frac13\,P_+^3 - \left(s^2-\mu\right)\,P_+\right] \nonumber \\
\left\{P_+(s),\,P_+(s')\right\}\propto \partial_s\,\delta(s-s'),
\end{eqnarray}
and is considered in \cite{Jevicki:1993qn},\cite{Polchinski:1994mb}. It describes
the effective field theory for the matrix quantum mechanics. As well this Hamiltonian describes
large wavelength KdV type (Burger--Hopf) equation with an external ($-s^2$) potential. It is known to be integrable. The relation of the matrix field theory to the string theory in a bit different setting was discussed as well in \cite{Akhmedov:2005mr}.

{\bf Conclusion.} We see that the same effective Das--Jevicki field theory appears  in the same matrix quantum mechanics (field theory in general) in two seemingly different approaches.
The present form of the Das--Jevicki theory follows from the Gaussian matrix quantum mechanics (field theory in general), while in our case we obtain it as the IR limit of the RG dynamics, where the matrix field theory under consideration flows to the Gaussian IR stable point.
Furthermore, the two ways of the derivation of this effective field theory should be related. We as well as Das and Jevicki, in fact, derive the same effective field theory on the same phase space --- VEVs of the operators and their sources in the matrix field theory. Note that the extra coordinate $s$ in the Hamiltonian has the same origin as in the paper \cite{Das:1990kaa}: there they have been using a bit different basis of operators $\Tr[e^{i\,k\,\phi}]$ from ours, which probably explains the reason why we did not get the Das--Jevicki Hamiltonian directly. The time in the Das--Jevicki theory appears to be one of the $x$'s of the matrix field theory under consideration. However, in the string field theory interpretation of the Das--Jevicki effective theory the time is related to the Liouville mode which defines the scale on the string world--sheet, while in our case it is just the energy scale.

One of the interesting questions is to understand the meaning of the solutions of the full Hamiltonian system in question from the point of view of the RG. Note that Wilsonian RG requires to define the value of the field $g$ (sources) at the initial value of ``time'' $\Lambda$, while the value of the the momentum $p$ should be fixed at the final ``time'' $\lambda$. At the same time we observe seemingly unexpected kind of RG dynamics because in the IR limit the RG flow goes in cycles described by the angle--action variables of the integrable system in question. The RG dynamics of such a type was predicted in \cite{Morozov:2003ik}.

{\bf Acknowledgments} We would like to thank A.Gerasimov, S.Apenko, A.Zabrodin, A.Rosly, A.Zotov, S.Kharchev, A.Morozov, A.Marshakov, L.Freidel and A.Mironov for valuable discussions. E.T.M. would like to specially thank M.Olshanetsky for sharing his ideas and for his interest to our work. The work of E.T.A. was done under the partial financial support by grants for the Leading Scientific Schools NSh-6260.2010.2 and RFBR 08-02-00661-a. The work of E.T.A. and E.T.M. was supported by Ministry of Education and Science of the Russian Federation under contract 02.740.11.0608

\end{document}